\documentclass[10pt, conference]{IEEEtran}

\usepackage[T1]{fontenc}
\usepackage{paralist}
\usepackage{cite}
\usepackage{amsmath,amssymb,amsfonts}
\usepackage{algorithmic}
\usepackage{graphicx}
\usepackage{textcomp}
\usepackage{xcolor}
\usepackage[inline]{enumitem} 
\usepackage{enumerate} 
\usepackage{multirow}
\usepackage{balance}
\usepackage{hyperref}
\usepackage[ruled,vlined,linesnumbered]{algorithm2e}
\usepackage{tablefootnote}
\usepackage{subcaption} 
\usepackage{dblfloatfix}

\SetCommentSty{mycommfont}

\usepackage{pifont}

\usepackage{xcolor}

\makeatletter
\IEEEtriggercmd{\reset@font\normalfont\footnotesize}
\makeatother
\IEEEtriggeratref{1}

\newcommand{\GGML}{\textit{GGML}\xspace}

\newcommand{\qki}{\textit{Q8\_K}\xspace}

\newcommand{\q}[1]{\textit{Q{#1}\_K}}

\newcommand{\LPP}{\textit{llama.cpp}\xspace}

\begin{document}
\bstctlcite{IEEEexample:BSTcontrol}

\title{F-BFQ: Flexible Block Floating-Point Quantization Accelerator for LLMs}

\author{Jude Haris, Jos\'e Cano \\
\emph{School of Computing Science, University of Glasgow, Scotland, UK}
}


\maketitle


\begin{abstract}

Large Language Models (LLMs) have become increasingly prominent for daily tasks, from improving sound-to-text translation to generating additional frames for the latest video games.
With the help of LLM inference frameworks, such as \LPP, which support optimizations such as KV-caching and quantization, it is now easier than ever to deploy LLMs on edge devices. 
Quantization is fundamental to enable LLMs on resource-constrained edge devices, and \LPP utilizes block floating point (BFP) quantization to drastically reduce the bit width of weights and input tensors, the memory footprint, and the computational power required to run LLMs.
LLMs are typically quantized with mixed BFP quantization across the model layers to reduce the loss of model accuracy due to quantization.
Therefore, to efficiently accelerate across the layers of BFP-quantized LLMs, specialized accelerators need to support different BFP variants without reconfiguration. 

To address this issue, we propose a Flexible Block Floating-Point Quantization (F-BFQ) accelerator, which can dynamically switch between two BFP quantization variants and perform matrix multiplication (MatMul) operations.
Our initial F-BFQ accelerator design, deployed on the AMD Kria board, reduces inference time by $1.4\times$ on average over the Arm NEON-based CPU execution across three BFP quantized LLMs while achieving $5.2$ tokens per second (${\sim}3.9$ words per second).




\end{abstract}


\section{Introduction}

Large language models (LLMs) such as the Llama~\cite{touvron2023llamaopenefficientfoundation} and GPT~\cite{Radford2019LanguageMA,brownLanguageModelsAre2020} family of models have revolutionized the ability of Artificial Intelligence (AI) systems to understand and generate human language in terms of text, audio, or video.
LLMs are an emerging class of machine learning (ML) models that are built by learning from huge text-based datasets.
With the innovation in model architecture and training methods, and through the help of the popularity of online services like ChatGPT~\cite{rayChatGPTComprehensiveReview2023}, the field of LLMs is evolving rapidly. 
The number of users is also growing rapidly due to the countless applications and use-cases from  classification~\cite{sunTextClassificationLarge2023}, code generation~\cite{liuYourCodeGenerated2023}, translation~\cite{yaoBenchmarkingLLMbasedMachine2024} to healthcare~\cite{clusmannFutureLandscapeLarge2023}.

Cloud-based LLM services like Gemini~\cite{google_gemini_overview_2024} have become the go-to method for daily users to access to LLMs. 
However, as the availability of open-source LLMs and datasets has increased, especially over the last few years, the need for edge-based, localized access and execution of LLMs has become more sought after due to concerns over security and data privacy.
The latest community-driven pushes have facilitated easy access to LLMs and rapid prototyping of new models and optimizations, enabling efficient LLM inference on edge devices.
The GPT-Generated Model Language~\cite{GGML} (GGML), a tensor library for ML specialized to enable high performance for LLMs on commodity hardware, is at the forefront of these pushes.
Furthermore, the \LPP inference framework~\cite{LLP}, which is based on the \GGML library, is specialized towards running LLMs on edge devices, supporting LLM inference on commodity CPUs, GPUs and NPUs.

Unfortunately, LLMs can be very computationally demanding, even for inference.
In addition, due to their large memory footprint, they require high memory capacity and bandwidth.
These properties of LLMs make them challenging to execute on resource-constrained edge devices.
For example, running LLMs on mobile phones or Internet-of-Things (IoT) devices is sometimes impossible due to memory constraints.
To improve performance at the edge, FPGA-based (Field Programmable Gate Arrays) accelerators~\cite{Khan2021NPE,Lu2020Transformer} have been developed to outperform standard CPU-based inference, but the problem of model size and inference memory footprint is still a limiting factor due to limited memory of the on-chip accelerator.

\begin{figure}[!t]
 \centering
 \includegraphics[width=0.85\columnwidth]{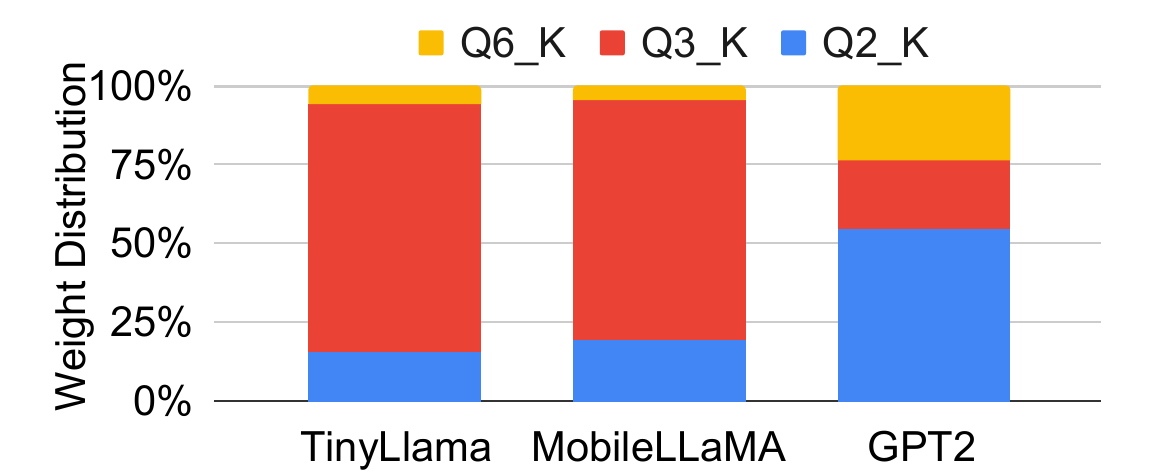}
  \caption{\label{fig:model_weight_breakdown} Weights quantization distribution percentage breakdown of the three LLMs under study.}
\end{figure}

Quantization is one of the key optimizations that is used to run LLMs on resource-constrained edge devices~\cite{gibsonDLASConceptualModel2025}.
Many quantization schemes have been developed and adopted to enable reduction of LLM sizes.
\LPP has specifically embraced the block floating point (BFP) quantization scheme, which enables blocks of data to be quantized in a group-like manner to further reduce the bit width.
To ensure minimal LLM accuracy reduction during quantization, statistical analysis is typically used to determine the level of quantization that can be applied without losing model accuracy.
To achieve the best trade-off between LLM accuracy and reduction in size, the level of quantization applied is chosen on a per-layer basis, depending on the weights of each layer.

Figure~\ref{fig:model_weight_breakdown} provides a breakdown of the distribution of weight parameters across three LLMs, GPT2~\cite{Radford2019LanguageMA}, MobileLLaMA~\cite{chu2023mobilevlmfaststrong} and TinyLlama~\cite{zhangTinyLlamaOpenSourceSmall2024} quantized using \LPP-based BFP quantization.
These models use three different variants of BFP quantization (\q{2}, \q{3}, and \q{6}) for the matrix multiplication (MatMul) operations.
Hence, to efficiently accelerate these quantized models, an accelerator that supports multiple BFP quantization variants is needed.

In this work, we propose a new Flexible Block Floating-Point Quantization (F-BFQ) accelerator that supports multiple BFP quantization variants to perform the MatMul operation with LLMs.
Our initial F-BFQ accelerator design focuses on supporting the \q{2} and \q{3} variants as a proof of concept.
We designed and evaluated our accelerator using the SECDA-LLM~\cite{haris2024designingefficientllmaccelerators} platform across three different LLMs.
The contributions of this work are as follows:

\begin{itemize}
    \item \textit{F-BFQ}, a new Flexible Block Floating-Point Quantization accelerator that supports \q{2} and \q{3} BFP MatMul operations.

    \item \textit{Dynamic Super-Block Vector Processor Unit}, a hardware module design that processes \q{2} and \q{3} operations concurrently, an is able to dynamically switch data loading and output accumulation depending on the required quantization variant per layer/operation.
    
    \item Evaluation of our initial accelerator across three LLMs (GPT2~\cite{Radford2019LanguageMA}, MobileLLaMA~\cite{chu2023mobilevlmfaststrong} and TinyLlama~\cite{zhangTinyLlamaOpenSourceSmall2024}), achieving an average speedup of $1.4\times$ over the baseline execution of the Arm NEON-based CPU with an average of $5.2$ tokens per second (${\sim}3.9$ words per second).
\end{itemize}

\section{Background and Related Work}


\subsection{Large Language Models}

Large Language Models (LLMs) are a class of machine learning models built upon the Transformer architecture~\cite{vaswani2017attention} and pre-trained on massive language corpora.
Typically, these models are adapted to specific downstream tasks, such as question-answering, via fine-tuning on task-relevant datasets.
LLMs are characterized by their large parameter counts which increases memory footprint; for instance, LLaMA models start with 7 billion parameters~\cite{touvron2023llama}.
Many LLMs operate in an auto-regressive manner, predicting the next tokens (or words) based on previously cached context.
This approach, known as key-value (KV) caching~\cite{kwon2023efficient}, enhances performance but incurs linearly scaling memory overhead.

Quantization techniques are widely adopted to reduce the parameter count and enable the deployment of LLMs on resource-constrained edge devices.
For example, 8-bit quantization has been shown to maintain accuracy while reducing model size by up to 4x~\cite{Zafrir2019Q8BERT, Wan2024ASP-DAC}.
Researchers have also explored more aggressive strategies, such as 4-bit quantization, to further shrink models without compromising accuracy~\cite{Shen2024EdgeQAT}.
In this work we look at a more promising approach, block floating point (BFP) quantization, which has been evaluated against traditional integer quantization methods~\cite{Rouhani2023Microscaling} for its potential advantages in efficiency and performance.


\subsection{BFP Quantization in \LPP}

\LPP~\cite{LLP} is a pure C/C++ library with minimal external dependencies for enabling LLMs inference on a wide range of edge devices that support GCC/Clang.
Currently, \LPP supports a wide set of LLMs, including some multi-modal and custom-defined models.
With the GPT-Generated Unified Format model format (GGUF) used by \LPP, it is possible to represent the weights of an LLM with as few as 1.5 bits using BFP quantization.
These quantized weights enable users to run LLMs on resource-constrained edge devices such as the Raspberry Pi and the Pixel phone~\cite{simonwillison}.

\begin{figure}[!t]
  \centering
  \includegraphics[width=0.85\columnwidth]{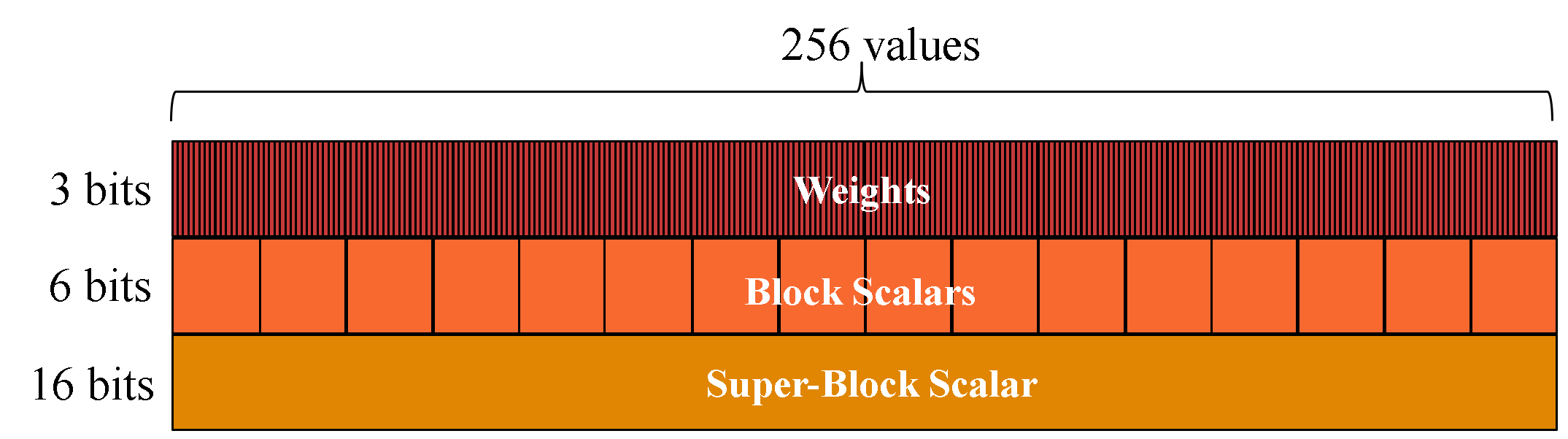}
  \caption[Q3\_K super-block Data Format]{\label{fig:q3_k_format} Q3\_K super-block Data Format.}
 \end{figure}

\LPP~\cite{LLP} supports BFP quantization variants of 1.5, 2, 3, 4, 5, 6, and 8 bits along with some additional quantization techniques to recover accuracy.
These variants are typically denoted as \( Q{x\_y} \), where \( x \) represents the number of bits per weight and \( y \) denotes the type of quantization.
For example, Figure~\ref{fig:q3_k_format} shows the \q{3}~\cite{gerganovGGUFQuantizations} BFP format, where a superblock (SB) represents 256 weights and is partitioned into 16 blocks.
Each block contains 16 weights and a block-scaling factor \textit{BSF} of 6-bits.
Each SB also has a super-scaling factor \textit{SSF} of 16 bits.
The \textit{BSF} and \textit{SSF} values are used to rescale the weights and ensure minimal accuracy loss.
By summing up the total number of bits required for the weights and the scaling factors and then dividing by the number of weights, we can determine that the BFP format requires $\sim$3.5 bits per weight, which is a significant reduction compared to the typical 32-bit floating-point format used in LLMs.
The \q{2}~\cite{gerganovGGUFQuantizations} format differs from \q{3}, as it contains 2-bit weights, 4-bit `minimum' and `scalar' values for each block (total 8-bits per block), and 16-bit `minimum' and `scalar' values for the superblock (total 32-bits per superblock). 
Overall, it requires around $\sim$2.6 bits per weight.
Note that the \qki format is used for input tensors, where each SB contains 256 input values and a single \textit{SSF} of 16 bits, thus requiring $\sim$8 bits per input.

\subsection{BFP Acceleration}

Some previous works focus on supporting and accelerating BFP quantization for DNNs.
FlexBlock~\cite{nohFlexBlockFlexibleDNN2023} proposes a new accelerator for DNN training with multiple BFP support.
By dynamically switching to lower BFP bit widths, it allows up to $16\times$ more multiple-accumulate (MAC) operations than a static bit-width solution.
In terms of inference, Bucket Getter~\cite{loBucketGetterBucketbased2023} is a novel architecture for DNN inference that utilizes multiple smaller accumulators within their custom floating point adder to support different scales of BFP quantization operations. 

F-BFQ focuses on LLM inference with the GGUF BFP format, which is widely used in the \LPP framework.


\section{F-BFQ Accelerator Architecture}

Utilizing our SECDA-LLM\cite{haris2024designingefficientllmaccelerators} platform we were able to quickly and efficiently design our Flexible Block Floating-Point Quantized (F-BFQ) accelerator architecture.
The goal of the initial design is to provide a template architecture which supports two variants of BFP quantization.

The rest of this section provides: 
\begin{enumerate*}[label=\roman*)]
    \item an overview of the F-BFQ Architecture;
    \item descriptions of the hardware modules;
    \item in-depth look into the dynamic super-block processor;
    \item and a discussion about the software driver, including the opcode generation and tiling.
\end{enumerate*}


\subsection{Overview}

\begin{figure}[!t]
 \centering
 \includegraphics[width=0.85\columnwidth]{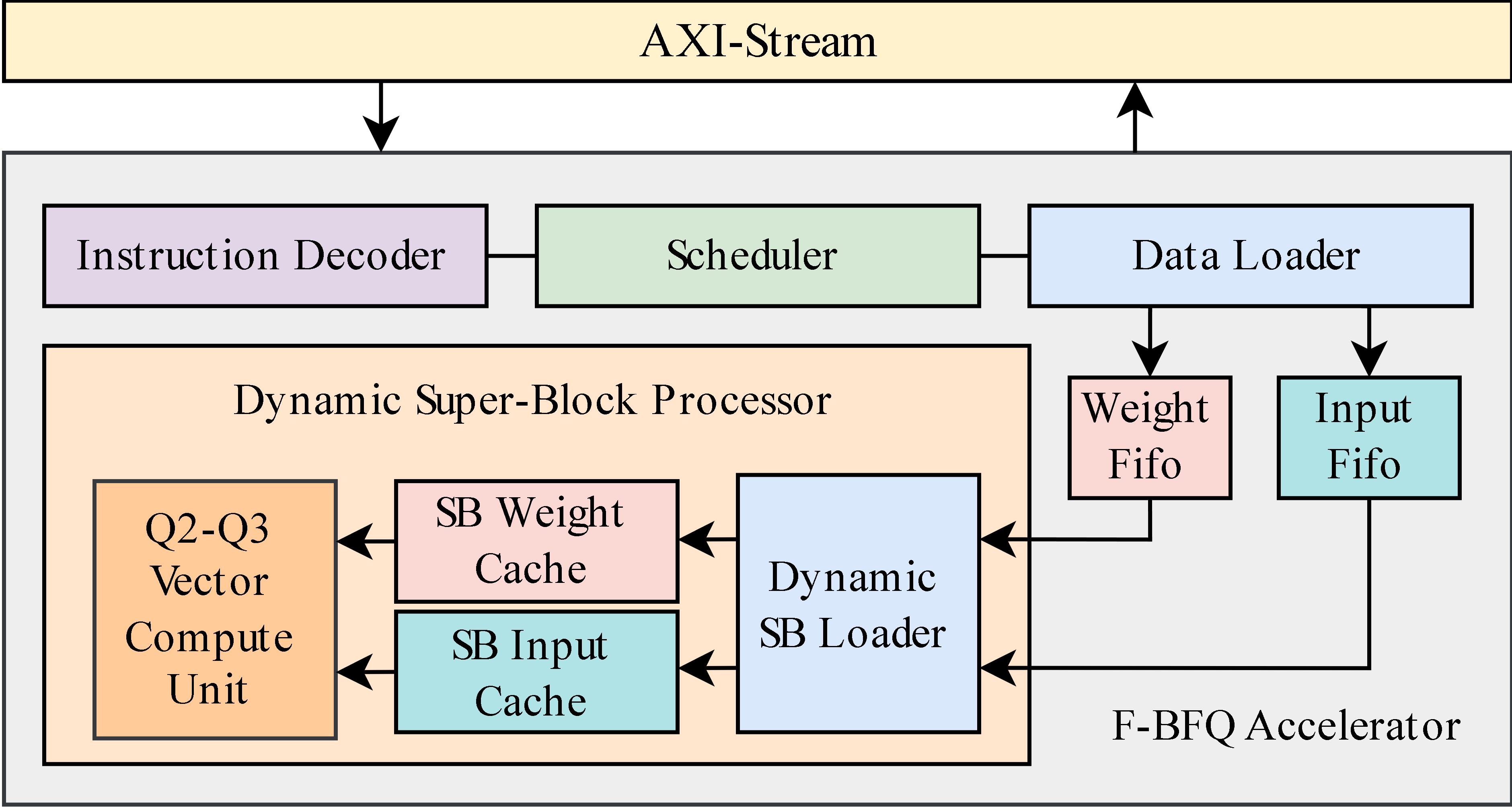} 
 \caption{\label{fig:llm_acc} Overview of our F-BFQ accelerator design for \q{2} and \q{3} MatMul operations.}
\end{figure}

Our F-BFQ accelerator design aimed to create a scalable architecture that could support efficient processing of BFP-quantized MatMul operations.
The proposed accelerator architecture, shown in Figure~\ref{fig:llm_acc}, contains an \emph{instruction decoder}, a \emph{data loader}, a \emph{scheduler} and the \emph{Dynamic Super-Block Processor (DSBP)}, along with data FIFOs that are used to temporary store weight and input data. The following sections describe the details of the main hardware components of the accelerator.

\subsubsection{Instruction Decoder}
loads and decodes opcode-based instructions from the AXI-Stream (the data interface to main memory) and then communicates the instruction throughout the rest of the accelerator.
Table~\ref{tab:opcodes} shows the micro-ISA opcode set for the accelerator.
The opcodes are generated by the accelerator driver and are sent via the AXI-stream.
Note that some opcodes are immediately followed by operand data, which the accelerator expects once the instruction is decoded. 
For example, after opcode `0x02', a set of weight SBs are expected and parsed by the data loader.

\begin{table} [t]
  \centering
\fontsize{7.5}{9}\selectfont  
  \caption{Micro-ISA Opcode Set.}
  \label{tab:opcodes}
  \begin{tabular}{|c|c|}
  \hline
  \textbf{Opcode} & \textbf{Description}   \\ \hline
  0x01            & Configure DSBP (Sets configuration registers)        \\ \hline
  0x02            & Load Weights (Configures Data Loader to load Weights)            \\ \hline
  0x04            & Load Input (Configures Data Loader to load Inputs)             \\ \hline
  0x08            & Schedule MatMul Operation (Activates DSBP)         \\ \hline
  0x10            & Store Output (Sends Output Data to Main Memory)           \\ \hline
  \end{tabular}
\end{table}

\subsubsection{Data Loader}
parses the incoming data stream and maps the weight and input SBs into their respective data FIFOs; the number of SBs loaded is configured with the `0x01' opcode.
It also partitions consecutive data elements in depth dimension across $N$ FIFOs, enabling parallel access to data so that the \emph{DSBP} can compute $N$ operations simultaneously without stalling the computation pipeline.

\subsubsection{Scheduler}
tiles the MatMul problem according to the dimensions of the target layer and controls the DSBP.
Additionally, it synchronizes and accumulates the output data produced by the DSBP and sends the results back to main memory via the AXI-Stream.


\subsection{Dynamic Super Block Processor}

\begin{figure}[!t]
 \centering
 \includegraphics[width=0.85\columnwidth]{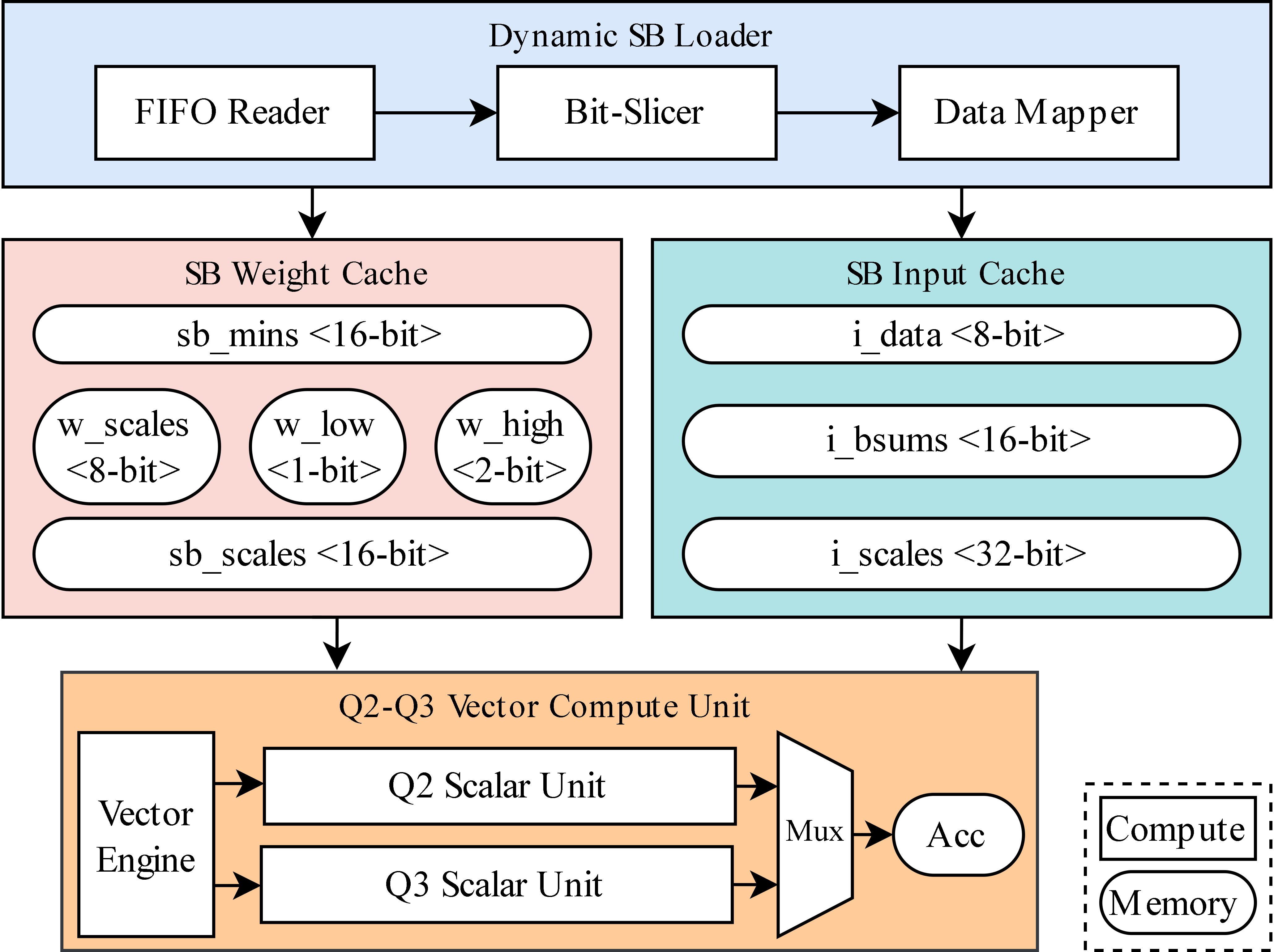} 
 \caption{\label{fig:dsbp} Detailed view of the Dynamic Super Block Processor.}
\end{figure}

The Dynamic Super Block Processor (DSBP) is the core processor within the F-BFQ accelerator.
While our current design contains a single DSBP, the F-BFQ accelerator is designed to be modular and scalable with multiple DSBPs.
The DSBP consists of four main components: Dynamic SB Loader, SB Weight Cache, SB Input Cache and the (Q2-Q3) Vector Compute Unit.
Figure~\ref{fig:dsbp} contains a detailed view of the DSBP architecture and its components.

The Dynamic SB Loader is responsible for loading data from the Input and Weight FIFOs into the respective local SB cache.
First, the `fifo reader' reads the data, then according to the type (input/weight) and the variant (\q{2},\q{3} or \q{8}), the `bit-slicer' slices the data packets into the different parameters, and finally the `data mapper' stores the data into the correct buffers within the SB caches.

For example, when \q{3} SB weight data is being read, the bit-slicer and the data mapper partition and store the SB data into `w\_{scales}',`w\_{low}' and the `w\_{high}' buffers within the SB weight cache according to the \q{3} SB data format shown in~\ref{fig:q3_k_format}.
The input and weight caches are stored in partitioned BRAM buffers, enabling the vector compute unit to access consecutive data in parallel.

The vector compute unit (VCU), in our case the Q2-Q3 VCU, supports both \q{2} and \q{3} SB vector operations.
To efficiently handle both variants while maintaining low resource overhead, the VCU contains a common `vector engine' that can perform the dot product operation required for matrix multiplication. 
The remaining scaling operations, which are dependent on the quantization variant, are handled by `Q2/Q3 Scalar Units'. 
Finally, a multiplexer (`Mux`) is used to accumulate the correct output value in the accumulator register (`Acc').
This accumulator register is then read and saved back to main memory.

\subsection{F-BFQ Driver}
\label{subsec:acc_driver}

The accelerator driver is a key component to enable seamless operation between the SECDA-LLM~\cite{haris2024designingefficientllmaccelerators} backend and the accelerator.
Our F-BFQ driver is configured to receive all MatMul operations that use \q{2} or \q{3} weights and \q{8} inputs. 
The driver is responsible for generating and sending opcode instructions through an AXI-Stream data transfer to control the accelerator: 
\begin{enumerate*}[label=\roman*)]
    \item First, the driver receives the necessary metadata to configure the accelerator from the SECDA-LLM backend; using this, it sends the `0x01' opcode to configure the DSBP with the corresponding dimension of the MatMul operation. Additionally, the driver updates the `weight\_{type}' control register which decides if the accelerator is in \q{2} or \q{3} mode;
    
    \item Second, the accelerator checks if the input matrix is small enough to be stored inside the input buffer without tiling; if so, the input matrix is sent to the accelerator. Otherwise, output stationary tiling is used to tile and send the weight and input data to the accelerator using the `0x02' and `0x04' opcodes, respectively. 
    Once the data blocks are sent to the accelerator, the '0x08' opcode is sent to the accelerator to start executing the MatMul operation;

    \item Finally, once the accelerator processes all the data, the `0x10' opcode is sent to the accelerator to transfer the output data to main memory.
\end{enumerate*}

\section{Evaluation}

To evaluate our F-BFQ accelerator design and understand the performance improvements, we fully utilize the integration with SECDA-LLM~\cite{haris2024designingefficientllmaccelerators} to perform end-to-end LLM inference.


\subsection{Experimental Setup}
\label{subsec:exp_setup}

We deploy and evaluate our accelerator design on the AMD KV260~\cite{amdKriaKV260} board.
Table~\ref{tab:device_stats} provides some hardware details along with the resource utilization of our accelerator design (in \%). 
We deploy three LLMs to evaluate our accelerator, and Table~\ref{tab:model_stats} provides key details about them, such as the number of MatMul layers, number of parameters, and model size.
These LLMs contain various levels of BFP quantization but always contain layers quantized to \q{3} and \q{2}.
We use the \textit{llama-cli} program from the \LPP framework to cross-compile for our targeted CPU architecture (ARMv8) with NEON vector instructions enabled alongside our accelerator driver.
Note that we run all our experiments $10$ times each and take the average to present the results.

\begin{table}
\caption{\label{tab:device_stats} Specifications of the KV260 board. We highlight the FPGA hardware resource used in terms of absolute value and \% of the total available. The FPGA runs at 200MHz.}
\centering
\fontsize{5.7}{9}\selectfont
\begin{tabular}{|c|c|c|c|c|c|c|}
\hline
\textbf{Device}   &\textbf{CPU}  &\textbf{DRAM } &\textbf{BRAM }   &\textbf{DSP}     &\textbf{FF}      &\textbf{LUT}      \\ \hline
KV260    &ARM-A53 &4GB    &234 (81\%)    &175 (14\%)    &14K (6\%)  &36K (30\%)     \\ \hline  
\end{tabular}
\end{table}

\begin{table}
\caption{\label{tab:model_stats} Specification of the LLMs used. Note the number of \q{2} and \q{3} MatMul layers within each model.}
\centering
\fontsize{6.3}{9}\selectfont
\begin{tabular}{|c|c|c|c|c|}
\hline

\textbf{Model}        & \textbf{\q{2} Layers}  & \textbf{\q{3} Layers}  & \textbf{Parameters} & \textbf{Size (MBs)} \\ \hline

GPT2~\cite{Radford2019LanguageMA}                   & 25 & 24                   & 163M         & 77             \\ \hline
TinyLlama~\cite{zhangTinyLlamaOpenSourceSmall2024}  & 45 & 110                  & 1.1B        & 460            \\ \hline
MobileLLaMA~\cite{chu2023mobilevlmfaststrong}             & 49 & 120                 & 1.4B       & 560           \\ \hline

\end{tabular}
\end{table}

\begin{table}[h]
\caption{\label{tab:results} Results for LLM inference using our F-BFQ accelerator design in terms of overall execution time (seconds), speedup compared to the CPU baseline, and tokens per second across the three LLMs under study.}
\centering
\fontsize{6.3}{9}\selectfont
\begin{tabular}{|c|c|c|c|c|}
\hline
\textbf{Model} & \textbf{Hardware} & \textbf{Overall (s)} & \textbf{Speedup} & \textbf{token/s} \\ \hline
\multirow{2}{*}{GPT2}         
            & CPU   & 1.85   & 1.00 & 8.31  \\ \cline{2-5}
            & FBFQ  & 1.58   & 1.17 & 12.18 \\ \hline
\multirow{2}{*}{MobileLLaMA}  
            & CPU   & 21.78  & 1.00 & 0.69  \\ \cline{2-5}
            & FBFQ  & 14.40  & 1.51 & 1.44  \\ \hline
\multirow{2}{*}{TinyLlama}    
            & CPU   & 17.59  & 1.00 & 0.86  \\ \cline{2-5}
            & FBFQ  & 11.49  & 1.53 & 1.82  \\ \hline
\end{tabular}
\end{table}


\subsection{Results}

We evaluated the accelerator performance in terms of inference time versus the CPU baseline on the KV260 board.
Table~\ref{tab:results} summarizes the results of these experiments.
The total time and speedup relative to the CPU available are presented in the `Overall' column.
Note that our input prompt passed to the LLMs contained $6$ tokens and that we requested each model to generate $10$ tokens during our experiments.
The `Overall' time represents the total time for both prompt processing and token generation; similarly, we consider both aspects for the `token/s' columns.

Overall, we achieve a speedup of $1.4\times$ on average across the three LLMs, reaching up to $12.2$ tokens/s for the GPT2 model.
We see that the GPT2 model, which is the smallest of the three, results in the lowest performance gains.
After further model analysis, which involved calculating the number of operations required for each layer, we conclude that this is due to the lower computational intensity of the model, especially during the token generation phase, which leads to high data bandwidth requirements.
We plan to support more quantization variants, such \q{4}-\q{8}, within F-BFQ and enable consecutive MatMul layers to be processed within the accelerator before sending the output data back to main memory to reduce the data transfer bandwidth bottleneck.


\section{Conclusion}

We proposed a new flexible block floating point (BFP) quantization accelerator, F-BFQ, for matrix multiplication operations within LLMs.
F-BFQ supports two BFP quantization variants by switching operational modes via opcodes generated by the F-BFQ driver.
We designed and implemented the accelerator architecture within the SECDA-LLM platform~\cite{haris2024designingefficientllmaccelerators} and evaluated it across three LLMs.
Compared to an Arm NEON-optimized CPU baseline, we obtained an average speedup of $1.4\times$, while averaging $5.2$ tokens per second.
Future work will support other BFP variants to accelerate all MatMul operations within a given BFP-quantized LLM.


\section*{Acknowledgment}

This work was supported by the EU Project dAIEDGE (GA Nr 101120726) and the Innovate UK Horizon Europe Guarantee (GA Nr 10090788).


\balance

\bibliographystyle{IEEEtran}
\bibliography{bib}

\end{document}